\documentclass[aps,prb,twocolumn,showpacs]{revtex4}
\usepackage{graphicx}
\usepackage{dcolumn}
\usepackage{bm}
\begin{document}
\title{Superconductivity of the Ternary Boride Li$_2$Pd$_3$B Probed by $^{11}$B NMR  }

\author{M. Nishiyama$^1$, Y. Inada$^2$ and Guo-qing~Zheng$^1$}
\affiliation{$^1$Department of Physics, Okayama University,  Okayama 700-8530,
Japan}

\affiliation{$^2$ Department of Science Education,
Faculty of Education, Okayama University,  Okayama 700-8530,
Japan}

Phys. Rev. {\bf B 71}, 220505(R) (2005)

\widetext
\begin{abstract}
We report a $^{11}$B NMR measurement on the recently discovered superconductor Li$_2$Pd$_3$B. The nuclear spin lattice relaxation rate $1/T_1$ shows a well-defined coherence peak just below $T_c$ ($H$=1.46 T)=5.7 K, and the spin susceptibility measured by the Knight shift also decreases below $T_c$. These results indicate that the superconductivity is of conventional nature, with an isotropic gap. Our results also suggest that the $p$-electrons of boron and the $d$-electrons of palladium that hybridize with boron $p$-electrons are primarily responsible for the superconductivity. 
  
\end{abstract}
\vspace*{5mm}

\maketitle
\sloppy

Since the discovery of high temperature superconductivity in copper oxides \cite{Bednorz}, compounds containing transition metal elements have become targets for searching new strongly-correlated  superconductors. In fact, superconductivity was discovered in the cobalt oxide Na$_{0.3}$CoO$_2 \cdot$1.3H$_2$O (Ref. \cite{Takada}), which was found to be of unconventional nature with strong electron correlations in the normal state \cite{Fujimoto}. Meanwhile, the discovery of superconductivity at 40 K in MgB$_2$ \cite{Akimitsu} has generated recurred interest on the physical properties of borides. Very recently, Togano {\it et al} found that the ternary metallic compounds containing boron and palladium, Li$_2$Pd$_3$B, is superconducting at $T_c \sim$ 7 K (Ref.\cite{Togano}). This compound has a cubic structure with the space group of P4$_3$32, containing a distorted Pd$_6$B octahedra, which is structurally similar to the superconductor MgCNi$_3$ ($T_c$=8 K) \cite{Cava} and in some sense also similar to the high-$T_c$ copper oxides where the key structure is the oxygen-containing octahedra. Although the physical properties of this new compound are unexplored, it has been proposed that the correlations of Pd $d$-electrons may be dominant in the electronic properties and may also be responsible for the superconductivity. \cite{Sardar}  Band calculation has shown that Pd $d$-electrons contribute significantly to the density of states at the Fermi level. \cite{Chandra} 
 
In this Communication, we report the first $^{11}$B nuclear magnetic resonance (NMR) measurement in the superconducting and the normal states. We find that the superconductivity is of Bardeen-Cooper-Schrieffer (BCS) type with an isotropic energy gap.  In the normal state, the temperature ($T$) dependencce of the nuclear spin-lattice relaxation rate $1/T_1$ and the  Knight shift due to spin susceptibility,  $K_s$ satisfies the so-called Korringa relation, $T_1TK_s^2$=const, with no signature of electron correlations. Our results suggest that the superconductivity in Li$_2$Pd$_3$B is phonon mediated.
 
The samples were prepared by the arc-melting method with starting materials of Li (99.9\% purity), Pd (99.95\%) and B (99.5\%).
The two-step arc melting process  \cite{Togano} was used.
The starting melt composition of Li:Pd:B=2.1:3:1 was adopted in the light of high vapor pressure of Li. 
 The x-ray diffraction chart shows that the sample is single phase. For NMR measurements, the sample was crushed into powder.
 Figure 1 shows the ac susceptibility measured using the NMR coil. The $T_c$ is about 7.1 K at zero magnetic field. 
 \begin{figure}
\begin{center}
\includegraphics[scale=0.5]{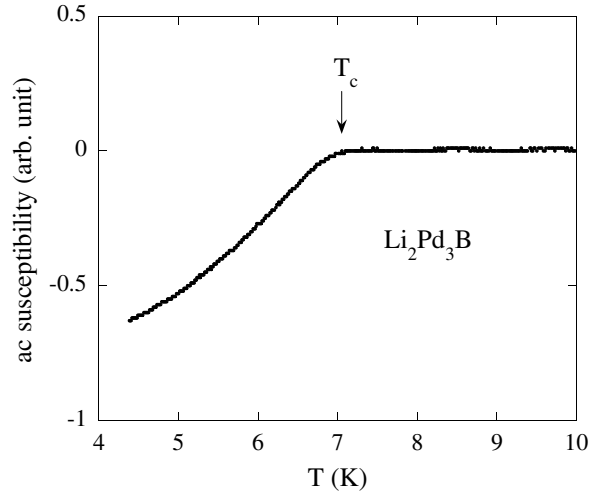}
\caption{ The ac susceptibility measured using the NMR coil for Li$_{2}$Pd$_{3}$B at zero magnetic field. }
\label{fig:1}
\end{center}
\end{figure}
 A standard phase-coherent pulsed NMR spectrometer was used to collect data. The NMR spectra were obtained by fast Fourier transform (FFT) of the spin echo taken at a constant magnetic field of 1.4629 T. The nuclear spin-lattice relaxation rate, $1/T_1$, was measured  by using a single saturation pulse and fitting the recovery of the nuclear magnetization  after the saturation pulse. 

Figure 2 shows the spectrum at $T$=10 K. 
The full width at the half maximum (FWHM) of the NMR line is less than 5 kHz. The very sharp transition ensures the high quality of the sample. The temperature dependence of the FWHM is shown in Figure 3. It can be seen that the spectrum broadens below $T_c(H$=1.46 T)=5.7 K, at which temperature both $1/T_1$ and the Knight shift shows anomaly (see below).

\begin{figure}
\begin{center}
\includegraphics[scale=0.5]{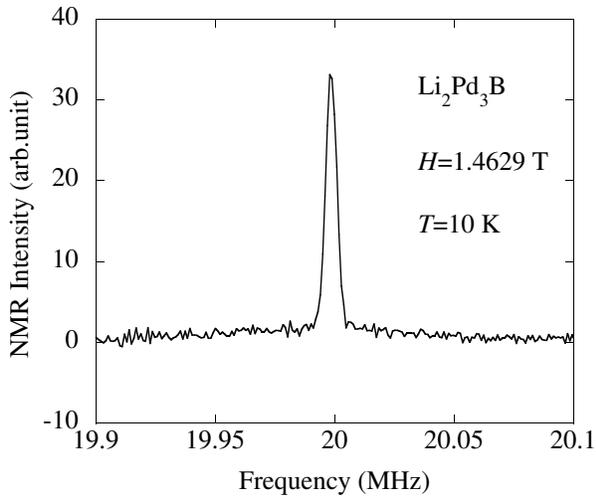}
\caption{ The  Fourier transformed $^{11}$B NMR spectrum measured at $T$=10 K at a magnetic field of $H$=1.4629 T. }
\label{fig:2}
\end{center}
\end{figure}

\begin{figure}
\begin{center}
\includegraphics[scale=0.5]{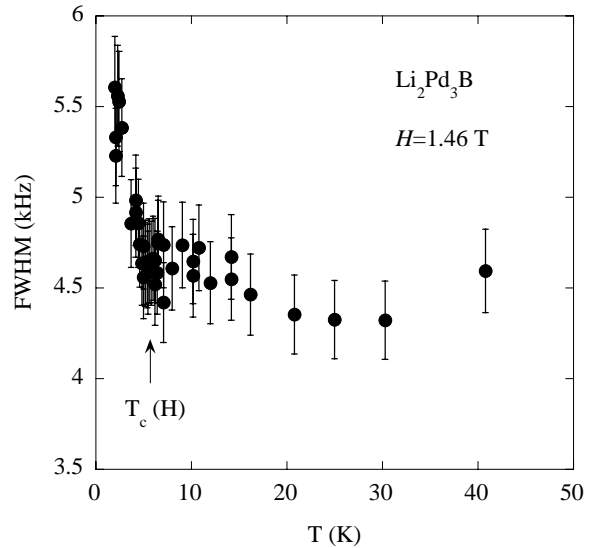}
\caption{ The temperature dependence of the  full width at the half maximum (FWHM) of the $^{11}$B NMR spectrum. }
\label{fig:3}
\end{center}
\end{figure}

Figure 4 shows the temperature dependence of $1/T_1$.  In the normal state above $T_c$, $1/T_1$ varies in proportion to $T$, as seen in conventional metals. Just below $T_c$ ($H$=1.46 T)=5.7 K, however, $1/T_1$ increases as $T$ is reduced, showing a coherence peak, then decreases exponentially upon further lowing $T$. This is the characteristic of  $s$-wave, isotropic superconductivity. The $1/T_{1s}$ in the superconducting state is expressed as 
\begin{eqnarray}
\frac{T_1(T=T_c)}{T_{1s}} & = & \nonumber\\
 \frac{2}{k_BT_c}\int (N_s(E)^{2}+M_s(E)^{2})f(E)(1-f(E))dE
\end{eqnarray}
where $N_{s}(E)=N_{0} E/(E^2-\Delta^2)^{1/2}$ is the superconducting density of states (DOS) with $\Delta$ being the BCS gap,  $M_{s}(E)=N_{0} \Delta/(E^2-\Delta^2)^{1/2}$ is the anomalous DOS due to the coherence factor \cite{Mac}, $N_{0}$ is the DOS in the normal state and
$f(E)$ is the Fermi function. Both $M_{s}(E)$ and $N_{s}(E)$ diverge at  $E=\Delta$, so theoretically $1/T_1$ diverges just below $T_c$. In real materials, however, broadening of the energy level and some anisotropy in the energy gap will remove the singularity at $E=\Delta$ so that $1/T_1$ only shows a peak just below $T_c$.
 Following Hebel \cite{Hebel}, we convolute $M_{s}(E)$ and $N_{s}(E)$ with a broadening function $B(E)$ which is  approximated with a rectangular function centered at $E$ with a height of $1/2\delta$. The solid curve below $T_c$ shown in Figure 4 is a calculation with 2$\Delta(0)=2.2k_BT_c$ and $r\equiv\Delta(0)/\delta$=2.  It fits the experimental data reasonably well. The parameter 2$\Delta(0)$ is substantially smaller than the BCS value of 3.5$k_BT_c(H)$. This could be due to the effect of the magnetic field, which usually reduces the gap size  \cite{Mac}.

\begin{figure}
\begin{center}
\includegraphics[scale=0.5]{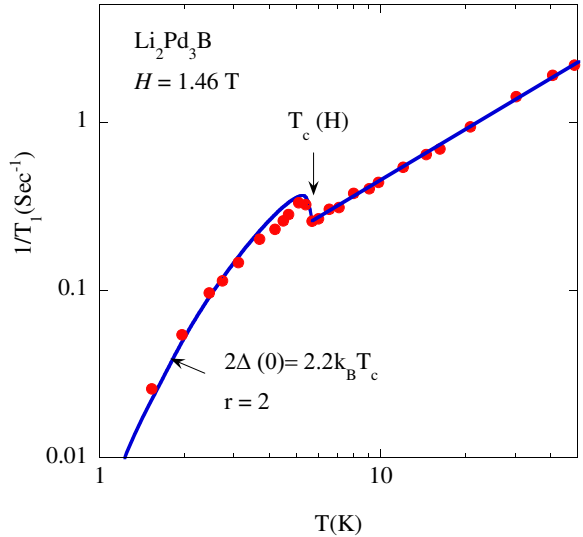}
\caption{(Color on-line) The  temperature dependence of the spin lattice relaxation rate, $1/T_1$, measured at a field of  $H$=1.4629 T. The straight line above $T_c$ represents the $T_1T$=const relation. The solid curve below $T_c$ is a calculation assuming the BCS gap function with the parameter shown in the figure. For detail, see the text. }
\label{fig:4}
\end{center}
\end{figure}

Figure 5 shows the Knight shift, $K$, as a function of temperature. 
The shift is small and  $T$-independent within the experimental error in the normal state, but changes abruptly at $T_c$ where $1/T_1$ shows the coherence peak. Note that the change is positive, namely, $K$ increases below $T_c$. 
The shift may be expressed as,
\begin{eqnarray}
K = K_{orb}+K_s
\end{eqnarray}
\begin{eqnarray}
K_s=A_{hf}\chi_s
\end{eqnarray}
where $K_{orb}$ is the contribution due to orbital (Van Vleck) susceptibility which is $T$-independent, $A_{hf}$ is the hyperfine coupling constant and $\chi_s$ is the spin susceptibility.
$K_s$ may  be decomposed into 
\begin{eqnarray}
K_s=K_{cp}+K_0
\end{eqnarray}
where $K_{cp}$ is the shift due to core polarization interaction between the boron $p$ electrons and the nuclear spins, which is negative, and $K_0$ is due to other interactions including the usual Fermi contact interaction. 
The result that $K_s$ increases below $T_c$ indicates that $K_{cp}$ is dominant over $K_0$. Therefore, the DOS that decreases due to the onset of superconductivity are due to boron $p$ electrons, and/or  $d$ electrons of palladium which hybridize with boron $p$ electrons. 
 If we assume that the spin susceptibility vanishes completely, due to singlet pairing, at $T$=1.6 K which is well below $T_c$, then $K_{orb}$ is about 0.085$\pm$0.003\% and $K_{s}$=-0.012$\pm$0.003\%. The curve below $T_c$ depicts the calculated Knight shift in the superconducting state 
 \begin{eqnarray}
K_s^{sc}\propto \chi_s^{sc}\propto - \int N_s(E)(\partial f(E)/\partial E)dE
\end{eqnarray}
 using the same gap parameter as in Figure 4. 
\begin{figure}
\begin{center}
\includegraphics[scale=0.5]{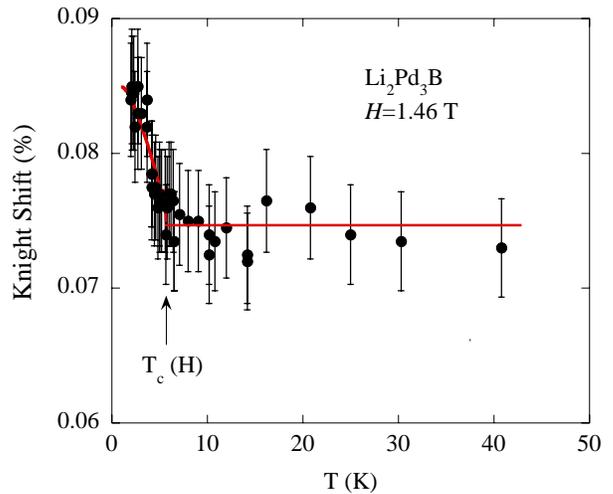}
\caption{ (Color on-line) The  temperature dependence of the Knight shift  measured at a constant field of  $H$=1.4629 T. The straight line above $T_c$ is a guide to the eyes. The curve below $T_c$ is the calculated result assuming a BCS gap with the same parameter as in the calculation of $T_1$.}
\label{fig:5}
\end{center}
\end{figure}

Finally, when the electrons  responsible for the spin lattice relaxation and the Knight shift are  $s$-electrons,   $T_1TK_s^2$=$\frac{\hbar}{4\pi k_B}(\frac{\gamma_e}{\gamma_N})^2$=2.55$\times$10$^{-6}$ Sec$\cdot$K (Korringa constant), whre $\gamma_e$ and $\gamma_N$ are the gyromagnetic ratio of electron and $^{11}$B nuclear spins, respectively. The observed value of  $T_1TK_s^2$ in the present case is 3.21$\times$10$^{-7}$ Sec$\cdot$K, which is much smaller than the Korringa constant. This result also supports that the dominant electrons that participate in the relaxation and the Knight shift are $p$ electrons, in which case the excess relaxation such as the orbital relaxation is important so that the observed $T_1TK_s^2$ deviates from $\frac{\hbar}{4\pi k_B}(\frac{\gamma_e}{\gamma_N})^2$.

In summary, through the measurements of $^{11}$B NMR, we find that the spin-lattice relaxation rate $1/T_1$ in the newly discovered Pd-containing boride Li$_2$Pd$_3$B exhibits a well-defined coherence peak just below $T_c$ and decreases exponentially with further decreasing temperature. The spin susceptibility as measured by the Knight shift $K_s$ also decreases below $T_c$, suggesting a singlet spin pairing. In the normal state, the Korringa relation, namely, $T_1TK_s^2$=const is obeyed, indicating the lack of electron correlations. Our results indicate that this compound is a  superconductor with an isotropic gap and   that the superconductivity occurs in the absence of electron correlations.

 We thank T. Kato and K. Matano for assistance in some of the measurements.  
This work was supported in part 
 by a Grant-in-Aid for
Scientific Research  from MEXT.

\end{document}